\documentclass[12pt,numbers,sort&compress]{article}

\usepackage{natbib}
\usepackage{hypernat}
\usepackage{amsmath}
\usepackage{amsthm}
\usepackage{amsfonts}          
\usepackage{amssymb}           
\usepackage[all]{xy}           
\usepackage{color}
\usepackage[backref,colorlinks=true,linktocpage]{hyperref}

\newlength{\xtrawidth}
\newlength{\xtraheight}
\def\Z{\mathbb{Z}}
\def\fnote#1#2{\begingroup\def\thefootnote{#1}\footnote{#2}
  \addtocounter{footnote}{-1}\endgroup}
\newcommand{\Rep}[1]{\ensuremath{\mathbf{#1}}}
\newcommand{\barRep}[1]{\ensuremath{\overline{\Rep{#1}}}}
\newcommand{\Vt}{{\ensuremath{\widetilde{V}}}}

\newcommand{\Xt}{{\ensuremath{\widetilde{X}}}}
\newcommand{\eref}[1]{eq.~\eqref{#1}}

\newcommand{\cref}[1]{Chapter~\ref{#1}}
\def\fnote#1#2{\begingroup\def\thefootnote{#1}\footnote{#2}
  \addtocounter{footnote}{-1}\endgroup}
\def\IP{\mathbb{P}}

\def\cO{{\mathcal O}}

\def\cF{{\mathcal F}}
\def\cFt{{\widetilde{\cF}}}

\def\tx{\Xt}
\def\dP9{{d\IP}_9}
\def\C{\mathbb{C}}

\def\atv{\wedge^2 \Vt}
\def\tv{\Vt}
\DeclareMathOperator{\ad}{ad}

\DeclareMathOperator{\Tr}{Tr}
\def\op1{{\mathcal O}_{\IP^1}}

\def\v12{V_1 \otimes V_2}
\def\b{\beta}

\newcommand{\sseq}[3]{0 \longrightarrow #1 \longrightarrow #2 \longrightarrow #3 \longrightarrow 0}
\def\z3z3{{\Z_3 \times \Z_3}}

\def\dirac{\slash{\! \! \! \! D}}

\newcommand{\dual}{\ensuremath{\vee}}

\DeclareMathOperator{\rank}{rank}

\DeclareMathOperator{\Span}{span}
\newcommand{\diff}{\mathrm{d}}

\definecolor{grey}{gray}{0.4}
\newcommand{\Hpq}[3]{\big({#2},{#3}\textcolor{grey}{\big|{#1}}\big)}
\newcommand{\Hst}[4]{\big[{#3},{#4}\textcolor{grey}{\big|{#2},{#1}}\big]}

\begin{document}
\begin{titlepage}

  \title{
    \hfill{\normalsize  hep-th/0510142} \\[1em]
    {\LARGE 
      Moduli Dependent $\mu$-Terms in 
      \\
      a Heterotic Standard Model
    }
    \author{Volker Braun$^{1,2}$, Yang-Hui He$^{1,3}$, Burt A.~Ovrut$^1$,
      and Tony Pantev$^{2}$
      \fnote{~}{vbraun,
        yanghe, ovrut@physics.upenn.edu;
        tpantev@math.upenn.edu}\\[0.5cm]
      {\normalsize $^1$
        Department of Physics,}
      {\normalsize $^2$
        Department of Mathematics,}\\
      {\normalsize University of Pennsylvania,} 
      {\normalsize Philadelphia, PA 19104--6395, USA}\\
      {\normalsize $^3$
        Merton College, University of Oxford, OX1 4JD, UK} \\      
    }
    \date{}
  }

  \maketitle
  \begin{abstract}

    In this paper, we present a formalism for computing the
    non-vanishing Higgs $\mu$-terms in a heterotic standard model.
    This is accomplished by calculating the cubic product of the
    cohomology groups associated with the vector bundle moduli
    ($\phi$), Higgs ($H$) and Higgs conjugate ($\bar{H}$) superfields.
    This leads to terms proportional to $\phi H \bar{H}$ in the low
    energy superpotential which, for non-zero moduli expectation
    values, generate moduli dependent $\mu$-terms of the form $\langle
    \phi\rangle H \bar{H}$.  It is found that these interactions are
    subject to two very restrictive selection rules, each arising from
    a Leray spectral sequence, which greatly reduce the number of
    moduli that can couple to Higgs--Higgs conjugate fields. We apply
    our formalism to a specific heterotic standard model vacuum. The
    non-vanishing cubic interactions $\phi H \bar{H}$ are explicitly
    computed in this context and shown to contain only four of the
    nineteen vector bundle moduli.
  
  \end{abstract}
  \thispagestyle{empty}
\end{titlepage}

\tableofcontents

\section{Introduction}

Obtaining non-vanishing Higgs $\mu$-terms, and setting the scale of
these interactions, is one of the most important issues in realistic
superstring model building~\cite{Binetruy:2005ez}. In this paper, we
present a formalism for computing these terms and explicitly
demonstrate, within an important class of $E_8 \times E_8$ superstring
vacua, that non-vanishing Higgs $\mu$-terms are generated in the low
energy effective theory. The scale of these $\mu$-terms is set by the
vacuum expectation values of a selected subset of vector bundle
moduli.

In a series of papers~\cite{HetSM1,HetSM2,HetSM3}, we presented a
class of ``heterotic standard model'' vacua within the context of the
$E_8 \times E_8$ heterotic superstring. The observable sector of a
heterotic standard model vacuum is $N=1$ supersymmetric and consists
of a stable, holomorphic vector bundle, $V$, with structure group
$SU(4)$ over an elliptically fibered Calabi-Yau threefold, $X$, with a
$\z3z3$ fundamental group. Each such bundle admits a gauge connection
which, in conjunction with a Wilson line, spontaneously breaks the
observable sector $E_8$ gauge group down to the $SU(3)_{C} \times
SU(2)_{L} \times U(1)_{Y}$ Standard Model group times an additional
gauged $U(1)_{B-L}$ symmetry. The spectrum arises as the cohomology of
the vector bundle $V$ and is found to be precisely that of the minimal
supersymmetric standard model (MSSM), with the exception of one
additional pair of Higgs--Higgs conjugate superfields. These vacua
contain \emph{no exotic multiplets} and exist for both weak and strong
string coupling.  All previous attempts to find realistic particle
physics vacua in superstring theories~\cite{Gross:1985rr, Sen:1985eb,
  Evans:1985vb, Breit:1985ud, Aspinwall:1987cn, Green:1987mn,
  Curio:1998vu, Andreas:1999ty, Donagi:1999gc, Krause:2000gp,
  Andreas:2003zb, Curio:2003ur, Curio:2004pf, Blumenhagen:2005ga,
  Blumenhagen:2005zg, Becker:2005sg, Raby:2005vc} have run into
difficulties. These include predicting extra vector-like pairs of
light fields, multiplets with exotic quantum numbers in the low energy
spectrum, enhanced gauge symmetries and so on. A heterotic standard
model avoids all of these problems.

Elliptically fibered Calabi-Yau threefolds with ${\mathbb{Z}}_{2}$ and
${\mathbb{Z}}_{2} \times {\mathbb{Z}}_{2}$ fundamental group were
first constructed in~\cite{SM-bundle1,SM-bundle2,SM-bundle3}
and~\cite{z2z2-1,z2z2-2} respectively.  More recently, the existence
of elliptic Calabi-Yau threefolds with $\z3z3$ fundamental group was
demonstrated and their classification given in~\cite{dP9Z3Z3}.
In~\cite{DonagiPrincipal, FMW1, FMW2, FMW3}, methods for building
stable, holomorphic vector bundles with arbitrary structure group in
$E_8$ over simply-connected elliptic Calabi-Yau threefolds were
introduced. These results were greatly expanded in a number of
papers~\cite{SM-bundle1, SM-bundle2, SM-bundle3, Diaconescu:1998kg,
  Donagi:2004qk, Donagi:2004ia} and then generalized to elliptically
fibered Calabi-Yau threefolds with non-trivial fundamental group
in~\cite{SM-bundle3,Ovrut:2002jk,z2z2-1,z2z2-2}. To obtain a realistic
spectrum, it was found necessary to introduce a new
method~\cite{SM-bundle1,SM-bundle2,SM-bundle3,z2z2-1,z2z2-2} for
constructing vector bundles. This method, which consists of building
the requisite bundles by ``extension'' from simpler, lower rank
bundles, was used for manifolds with ${\mathbb{Z}}_{2}$ fundamental
group in~\cite{MR1797016, MR1807601, SM-bundle3, SU5-z2-1, SU5-z2-2}
and in the heterotic standard model context in~\cite{dP9Z3Z3}.
In~\cite{HetSM1,HetSM2,HetSM3,SU5-z2-1,SU5-z2-2}, it was shown that to
compute the complete low-energy spectrum of such vacua one must 1)
evaluate the relevant sheaf cohomologies, 2) find the action of the
finite fundamental group on these spaces and, finally, 3) tensor this
with the action of the Wilson line on the associated representation.
The low energy spectrum is the invariant cohomology subspaces under
the resulting group action. This method was applied
in~\cite{HetSM1,HetSM2,HetSM3} to compute the exact spectrum of all
multiplets transforming non-trivially under the action of the low
energy gauge group. The accompanying natural method of
``doublet-triplet'' splitting was also discussed.  In a recent
paper~\cite{ModHetSM}, a formalism was presented that allows one to
enumerate and describe the multiplets transforming trivially under the
low energy gauge group, namely, the vector bundle moduli.

Using the above work, one can construct a heterotic standard model and
compute its entire low-energy spectrum. As mentioned previously, the
observable sector spectrum is very realistic, consisting exclusively
of the three chiral families of quarks/leptons (each family with a
right-handed neutrino), two pairs of Higgs--Higgs conjugate fields and
a small number of uncharged geometric and vector bundle moduli.
However, finding a realistic spectrum is far from the end of the
story. To demonstrate that the particle physics in these vacua is
realistic, one must construct the exact interactions of these fields
in the effective low energy Lagrangian. These interactions occur as
two distinct types. Recall that the matter part of an $N=1$
supersymmetric Lagrangian is completely described in terms of two
functions, the superpotential and the Kahler potential. Of these, the
superpotential, being a ``holomorphic'' function of chiral
superfields, is much more amenable to computation using methods of
algebraic geometry. The terms of the superpotential itself break into
several different types, such as Higgs $\mu$-terms and Yukawa
couplings. In this paper, we begin our study of holomorphic
interactions by presenting a formalism for computing Higgs
$\mu$-terms. We apply this method to calculate the non-vanishing
$\mu$-terms in a heterotic standard model.

Specifically, we do the following. In Section~\ref{sec:prelim}, we
review the relevant facts about the structure of heterotic standard
model vacua and present the explicit example which we are going to
use. The formalism for computing the low energy spectrum is briefly
discussed and we give the results for our explicit choice of heterotic
standard model vacuum.  For example, the spectrum contains nineteen
vector bundle moduli.  Higgs $\mu$-terms are then analyzed and shown
to occur as the triple product involving two cohomology groups, one
giving rise to vector bundle moduli ($\phi$) and the other to Higgs
($H$) and Higgs conjugate ($\bar{H}$) fields in the effective low
energy theory.  For non-vanishing moduli expectation values, Higgs
$\mu$-terms of the form $\langle \phi \rangle H \bar{H}$ are generated
in the superpotential. Section~\ref{sec:LSS1} is devoted to discussing
the first Leray spectral sequence, which is associated with the
projection of the covering threefold $\tx$ onto the base space $B_2$.
The Leray decomposition of a sheaf cohomology group into $(p,q)$
subspaces is discussed and applied to the cohomology spaces relevant
to Higgs $\mu$-terms. It is shown that the triple product is subject
to a $(p,q)$ selection rule which severely restricts the allowed
non-vanishing terms. Specifically, we find that only four out of the
nineteen vector bundle moduli can participate in Higgs $\mu$-terms.
The second Leray decomposition, associated with the projection of the
space $B_2$ onto its base $\IP^1$, is presented in
Section~\ref{sec:LSS2}. The decomposition of any cohomology space into
its $[s,t]$ subspaces is discussed and applied to cohomologies
relevant to Higgs $\mu$-terms. We show that $\mu$-terms are subject to
yet another selection rule associated with the $[s,t]$ decomposition.
Finally, it is demonstrated that the subspaces of cohomology that form
non-vanishing cubic terms project non-trivially onto moduli, Higgs and
Higgs conjugate fields under the action of the $\z3z3$ group. This
demonstrates that non-vanishing moduli dependent Higgs $\mu$-terms
proportional to $\langle \phi \rangle H \bar{H}$ appear in the low
energy superpotential of a heterotic standard model.

Other holomorphic interactions in the superpotential, such as Yukawa
couplings and moduli dependent``$\mu$-terms'' for possible exotic
vector-like multiplets will be presented in up-coming publications.
The more difficult issue of calculating the K{\"a}hler potentials in a
heterotic standard model will be discussed elsewhere.

\section{Preliminaries}
\label{sec:prelim}

\subsection{Heterotic String on a Calabi-Yau Manifold}
\label{sec:CY}

The observable sector of an $E_8 \times E_8$ heterotic standard model
vacuum consist of a stable, holomorphic vector bundle, $V$, with
structure group $SU(4)$ over a Calabi-Yau threefold, $X$, with
fundamental group $\z3z3$.  Additionally, the vacuum has a Wilson
line, $W$, with $\z3z3$ holonomy. The $SU(4)$ instanton breaks the low
energy gauge group down to its commutant,
\begin{equation} 
  \xymatrix@C=15mm{
    E_8 \ar[r]^-{SU(4)} & Spin(10)
  }.
  \label{1}
\end{equation} The $Spin(10)$ group is then spontaneously broken by
the holonomy group of $W$ to
\begin{equation} 
  \xymatrix@C=15mm{
    Spin(10) 
    \ar[r]^-{\z3z3} &
    SU(3)_{C} \times SU(2)_{L} \times U(1)_{Y} \times U(1)_{B-L}
  }.
  \label{2}
\end{equation} In this way, the standard model gauge group emerges in
the low energy effective theory multiplied by an additional $U(1)$
gauge group whose charges correspond to $B-L$ quantum numbers.

The physical properties of this vacuum are most easily deduced not
from $X$ and $V$ but, rather, from two closely related entities, which
we denote by $\Xt$ and $\Vt$ respectively. $\Xt$ is
a simply-connected Calabi-Yau threefold which admits a freely acting
$\z3z3$ group action such that
\begin{equation} 
  X= \Xt\Big/\big(\z3z3\big).
  \label{add1}
\end{equation}
That is, $\Xt$ is a covering space of $X$. Similarly, $\Vt$ is a
stable, holomorphic vector bundle with structure group $SU(4)$ over
$\Xt$ which is equivariant under the action of $\z3z3$. Then,
\begin{equation} 
  V=\Vt\Big/\big(\z3z3\big).
  \label{add2}
\end{equation}
The covering space $\tx$ for a heterotic standard model was discussed
in detail in~\cite{dP9Z3Z3}.  Here, it suffices to recall that $\tx$
is a fiber product
\begin{equation} 
  \tx=B_1 \times_{\IP^1} B_2
  \label{3}
\end{equation}
of two special $\dP9$ surfaces $B_1$ and $B_2$. Thus, $\Xt$ is
elliptically fibered over both surfaces with the projections
\begin{equation} 
  \pi_1 : \tx \to B_1 \,, \quad \pi_2 : \tx \to B_2 \,.
  \label{4}
\end{equation}
The surfaces $B_1$ and $B_2$ are themselves elliptically fibered over
$\IP^1$ with maps
\begin{equation}
  \label{5} 
  \b_1 : B_1 \to \IP^1 \,, \quad \b_2 : B_2 \to \IP^1 \,.
\end{equation}
Together, these projections yield the commutative diagram
\begin{equation}
  \label{6} 
  \vcenter{\xymatrix@!0@=12mm{ & \Xt \ar[dr]^{\pi_2}
      \ar[dl]_{\pi_1} \\ B_1 \ar[dr]_{\beta_1} & & B_2 \ar[dl]^{\beta_2} \\
      & \IP^1 \,.  
    }}
\end{equation} 
The invariant homology ring of each special $\dP9$ surface is
generated by two $\z3z3$ invariant curve classes $f$ and $t$. Using
the projections in eq.~\eqref{4}, these can be lifted to divisor
classes
\begin{equation} 
  \tau_1 = \pi_1^{-1}(t_{1}) 
  \,, \quad 
  \tau_2 = \pi_2^{-1}(t_{2}) 
  \,, \quad 
  \phi = \pi_1^{-1}(f_{1}) = \pi_2^{-1}(f_{2})
  \label{7}
\end{equation}
on $\Xt$. These three classes generate the invariant homology ring of
$\Xt$.

\subsection{The Gauge Bundle}
\label{sec:bundle}

The crucial ingredient in a heterotic standard model is the choice of
the vector bundle $\Vt$. These bundles are constructed using a
generalization of the method of \emph{bundle
  extensions}~\cite{SM-bundle3,z2z2-2}.  Specifically, $\Vt$ is the
extension
\begin{equation}
  \label{8} 
  \sseq{V_2}{\Vt}{V_1}
\end{equation}
of two rank two bundles $V_1$ and $V_2$ on $\Xt$. A solution for
$V_{1}$ and $V_{2}$ for which $\Vt$ satisfies all physical constraints
was given in~\cite{HetSM3}.  The result is that
\begin{equation}
  \label{9}
  \begin{split} 
    V_1 =&~ \chi_2 \cO_{\Xt}(-\tau_1 + \tau_2) \oplus
    \chi_2 \cO_{\Xt}(-\tau_1 + \tau_2) 
    \\ 
    V_2 =&~ \cO_{\Xt}(\tau_1 - \tau_2) \otimes \pi_2^\ast W_{2} 
    ,
  \end{split}
\end{equation}
where $W_{2}$ is an equivariant bundle in the extension space of
\begin{equation}
  \sseq{\cO_{B_2}(-2f_{2})}{W_{2}}{\chi_2\cO_{B_2}(2f_{2}) \otimes I_9}
  \label{10}
\end{equation}
and for the ideal sheaf $I_9$ of 9 points we take a generic $\z3z3$
orbit. Here, $\chi_{2}$ is one of the two natural one-dimensional
representations of $\z3z3$ defined by
\begin{equation} 
  {\chi}_{1}(g_{1})=\omega \,, \quad
  {\chi}_{1}(g_{2})=1 \,; \qquad {\chi}_{2}(g_{1})=1 \,, \quad
  {\chi}_{2}(g_{2})=\omega \,,
  \label{11}
\end{equation}
where $g_{1,2}$ are the generators of the two $\Z_3$
factors, ${\chi}_{1,2}$ are two group characters of $\z3z3$, and
$\omega = e^{\frac{2 \pi i}{3}}$ is a third root of unity.  The
observable sector bundle $\Vt$ is then an equivariant element of the
space of extensions defined in eq.~\eqref{8}.

Let $R$ be any representation of $Spin(10)$ and $U(\Vt)_{R}$ the
associated tensor product bundle of $\Vt$.  Then, each sheaf
cohomology space $H^{i}(\Xt, U(\Vt)_{R})$, $i=0,1,2,3$
carries a specific representation of $\z3z3$. Similarly, the Wilson
line $W$ manifests itself as a $\z3z3$ group action on each
representation $R$ of $Spin(10)$.  As discussed in detail
in~\cite{HetSM3}, the low-energy particle spectrum is given by
\begin{multline}
  \label{12} 
  \ker\big(\dirac_{\Vt}\big) = \left( 
    H^0\big(\tx, \cO_{\tx}\big) \otimes \Rep{45} \right)^{\z3z3}
  \oplus 
  \left( H^1\big(\tx, \ad(\tv) \big) \otimes
    \Rep{1} \right)^{\z3z3} \oplus \\ \oplus 
  \left( H^1\big(\tx, \tv\big) \otimes
    \Rep{16} \right)^{\z3z3} \oplus 
  \left( H^1\big(\tx, \tv^\dual \big) \otimes
    \barRep{16} \right)^{\z3z3} \oplus 
  \left( H^1\big(\tx, \atv \big) \otimes \Rep{10} \right)^{\z3z3}
  ,
\end{multline} where the superscript indicates the invariant subspace
under the action of $\z3z3$.  The invariant cohomology space $(
H^0(\tx, \cO_{\tx}) \otimes \Rep{45} )^{\z3z3}$ corresponds to gauge
superfields in the low-energy spectrum carrying the adjoint
representation of $SU(3)_{C} \times SU(2)_{L} \times U(1)_{Y} \times
U(1)_{B-L}$. The matter cohomology spaces, $( H^1(\tx, \tv) \otimes
\Rep{16} )^{\z3z3}$, $( H^1(\tx, \tv^\dual ) \otimes \barRep{16}
)^{\z3z3}$ and $( H^1(\tx, \atv ) \otimes \Rep{10} )^{\z3z3}$ were all
explicitly computed in~\cite{HetSM3}, leading to three chiral families
of quarks/leptons (each family with a right-handed
neutrino~\cite{Giedt:2005vx}), no exotic superfields and two
vector-like pairs of Higgs--Higgs conjugate superfields respectively.
The remaining cohomology space, $(H^1(\tx, \ad(\tv)) \otimes
\Rep{1})^{\z3z3}$, was recently computed in~\cite{ModHetSM} and
corresponds to nineteen vector bundle moduli.

\subsection{Cubic Terms in the Superpotential}
\label{sec:couplings}

In this paper, we will focus on computing Higgs--Higgs conjugate
$\mu$-terms.  First, note that in a heterotic standard model Higgs
fields arise from eq.~\eqref{12} as zero modes of the Dirac operator.
Hence, they cannot have a ``bare'' $\mu$-term in the superpotential
proportional to $H \bar{H}$ with a constant coefficient. However,
group theory does allow $H$ and $\bar{H}$ to have cubic interactions
with the vector bundle moduli of the form $\phi H \bar{H}$.  If the
moduli develop a non-vanishing vacuum expectation value, then these
cubic interactions generate $\mu$-terms of the form $\langle \phi
\rangle H \bar{H}$ in the superpotential.  Hence, in a heterotic
standard model we expect Higgs $\mu$-terms that are linearly dependent
on the vector bundle moduli. Classically, no higher dimensional
coupling of moduli to $H$ and $\bar{H}$ is allowed.

It follows from eq.~\eqref{12} that the $4$-dimensional Higgs and
moduli fields correspond to certain $\bar{\partial}$-closed
$(0,1)$-forms on $\Xt$ with values in the vector bundle $\wedge^2 \Vt$
and $\ad(\Vt)$ respectively. Denote these forms by $\Psi_H$,
$\Psi_{\bar{H}}$, and $\Psi_\phi$. They can be written as
\begin{equation}
  \Psi_H = \psi^{(H)}_{\bar{\iota}, [ab]} 
  \,\diff \bar{z}^{\bar{\iota}}
  , \qquad
  \Psi_{\bar{H}} = \psi^{(\bar{H}),[ab]}_{\bar{\iota}}
  \,\diff \bar{z}^{\bar{\iota}}
  , \qquad
  \Psi_\phi = [\psi^{(\phi)}_{\bar{\iota}}]^{~b}_a
  \,\diff \bar{z}^{\bar{\iota}}
  , \qquad
\end{equation}
where $a$, $b$ are valued in the $SU(4)$ bundle $\Vt$ and
$\{z^\iota,\bar{z}^{\bar{\iota}}\}$ are coordinates on the Calabi-Yau
threefold $\Xt$. Doing the dimensional reduction of the
$10$-dimensional Lagrangian yields cubic terms in the superpotential
of the $4$-dimensional effective action. It turns out,
see~\cite{Green:1987mn}, that the coefficient of the cubic coupling
$\phi H \bar{H}$ is simply the unique way to obtain a number out of
the forms $\Psi_H$, $\Psi_{\bar{H}}$, $\Psi_\phi$. That is,
\begin{equation}
  W = \cdots + \lambda \phi H \bar{H}  
\end{equation}
where
\begin{equation}
  \label{eq:lambdaintegral}
  \begin{split}
    \lambda ~&= 
      \int_\Xt
      \Omega \wedge 
      \Tr\Big[ 
      \Psi_\phi \wedge \Psi_H \wedge \Psi_{\bar{H}} 
      \Big]
    = \\ &= 
      \int_\Xt
      \Omega \wedge 
      \Big( 
      [\psi^{(\phi)}_{\bar{\iota}}]^{~b}_a 
      \,
      \psi^{(H)}_{\bar{\kappa}, [bc]} 
      \,
      \psi^{(\bar{H}),[ca]}_{\bar{\lambda}}
      \Big)
      \diff \bar{z}^{\bar{\iota}} \wedge
      \diff \bar{z}^{\bar{\kappa}} \wedge
      \diff \bar{z}^{\bar{\lambda}}
  \end{split}
\end{equation}
and $\Omega$ is the holomorphic $(3,0)$-form. Mathematically, we are
using the wedge product together with a contraction of the vector
bundle indices to obtain a product
\begin{multline}
  \label{eq:cup}
  H^1\Big(\tx, \ad(\tv) \Big) \otimes 
  H^1\Big(\tx, \atv \Big) \otimes
  H^1\Big(\tx, \atv \Big) 
  \longrightarrow \\ \longrightarrow
  H^3\Big(\tx, \ad(\tv) \otimes \atv \otimes \atv \Big)
  \longrightarrow 
  H^3\Big(\tx, \cO_{\tx} \Big)  
\end{multline}
plus the fact that on the Calabi-Yau manifold $\Xt$
\begin{equation}
  H^3\Big(\Xt,\cO_{\Xt}\Big) =
  H^3\Big(\Xt, K_\Xt\Big) = 
  H_{\bar{\partial}}^{3,3}\Big(\Xt\Big) =
  H^6\Big(\Xt\Big)
\end{equation}
can be integrated over. If one were to use the heterotic string with
the ``standard embedding'', then the above product would simplify
further to the intersection of certain cycles in the Calabi-Yau
threefold. However, in our case there is no such description.

Hence, to compute $\mu$-terms, we must first analyze the cohomology
groups
\begin{equation} 
  H^1\Big(\tx, \ad(\tv)\Big),~
  H^1\Big(\tx, \atv \Big),~
  H^3\Big(\tx, \cO_{\tx} \Big)
  \label{14}
\end{equation}
and the action of $\z3z3$ on these spaces. We then have to evaluate
the product in eq.~\eqref{eq:cup}. As we will see in the following
sections, the two independent elliptic fibrations of $\Xt$ will force
most, but not all, products to vanish.

\section{The First Elliptic Fibration}
\label{sec:LSS1}

As discussed in detail in~\cite{HetSM3}, the cohomology spaces on
$\tx$ are obtained by using two Leray spectral sequences.  In this
section, we consider the first of these sequences corresponding to the
projection
\begin{equation} 
  \tx \stackrel{\pi_{2}}{\longrightarrow} B_2.
  \label{15}
\end{equation}
For any sheaf $\cal{F}$ on $\tx$, the Leray spectral sequence tells us
that\footnote{In all the spectral sequences we are considering in this
  paper, higher differentials vanish trivially. Hence, the $E_2$ and
  $E_\infty$ tableaux are equal and we will not distinguish them in
  the following. Furthermore, there are no extension ambiguities for
  $\C$-vector spaces.}
\begin{equation} 
  H^i\Big( \tx, {\cal F} \Big) = \bigoplus^{p+q=i}_{p,q}
  H^p\Big(B_2, R^q\pi_{2\ast}\cF\Big),
  \label{16}
\end{equation}
where the only non-vanishing entries are for $p=0,1,2$ (since $\dim_\C
(B_2)=2$) and $q=0,1$ (since the fiber of $\tx$ is an elliptic curve,
therefore of complex dimension one). Note that the cohomologies
$H^{p}(B_2, R^q\pi_{2\ast}\cF)$ fill out the $2\times 3$
tableau\footnote{Recall that the zero-th derived push-down is just the
  ordinary push-down, $R^0\pi_{2\ast}=\pi_{2\ast}$.}
\begin{equation}
  \label{17} 
  \vcenter{ \def\w{34mm} \def\h{8mm} \xymatrix@C=0mm@R=0mm{
      {\scriptstyle q=1} & 
      *=<\w,\h>[F]{ H^0\big(B_2, R^1\pi_{2\ast}\cF\big) } &
      *=<\w,\h>[F]{ H^1\big(B_2, R^1\pi_{2\ast}\cF\big) } & 
      *=<\w,\h>[F]{ H^2\big(B_2, R^1\pi_{2\ast}\cF\big) } 
      \\ {\scriptstyle q=0} & 
      *=<\w,\h>[F]{ H^0\big(B_2, \pi_{2\ast}\cF\big) } & 
      *=<\w,\h>[F]{ H^1\big(B_2, \pi_{2\ast}\cF\big) } & 
      *=<\w,\h>[F]{ H^2\big(B_2, \pi_{2\ast}\cF\big) } 
      \\ & {\scriptstyle p=0} & {\scriptstyle p=1} & 
      {\scriptstyle p=2} 
    }} 
  \Rightarrow 
  H^{p+q}\Big(\Xt, \cF\Big)
  ,
\end{equation}
where ``$\Rightarrow H^{p+q}\big(\Xt, \cF\big)$'' reminds us which
cohomology group the tableau is computing. Such tableaux are very
useful in keeping track of the elements of Leray spectral sequences.
As is clear from \eref{16}, the sum over the diagonals yields the
desired cohomology of $\cF$. In the following, it will be very helpful
to define
\begin{equation} 
  H^p\Big(B_2, R^q\pi_{2\ast}\cF\Big) \equiv 
  \Hpq{\cF}{p}{q}
  .
  \label{18}
\end{equation}
Using this notation, the tableau eq.~\eqref{17}. Using this
abbreviation, the tableau simplifies to
\begin{equation}
  \label{eq:Hpqtableau} 
  \vcenter{ \def\w{20mm} \def\h{8mm} \xymatrix@C=0mm@R=0mm{
      {\scriptstyle q=1} & 
      *=<\w,\h>[F]{ \Hpq{\cF}{0}{1} } &
      *=<\w,\h>[F]{ \Hpq{\cF}{1}{1} } &
      *=<\w,\h>[F]{ \Hpq{\cF}{2}{1} } 
      \\ {\scriptstyle q=0} & 
      *=<\w,\h>[F]{ \Hpq{\cF}{0}{0} } &
      *=<\w,\h>[F]{ \Hpq{\cF}{1}{0} } &
      *=<\w,\h>[F]{ \Hpq{\cF}{2}{0} } 
      \\ & {\scriptstyle p=0} & {\scriptstyle p=1} & 
      {\scriptstyle p=2} 
    }} 
  \Rightarrow 
  H^{p+q}\Big(\Xt, \cF\Big)
  .
\end{equation}


\subsection{The First Leray Decomposition of the Volume Form}

Let us first discuss the $(p,q)$ Leray tableau for the sheaf
$\cF=\cO_{\tx}$, which is the last term in eq.~\eqref{14}. Since the
sheaf is trivial, it immediately follows that
\begin{equation}
  \label{19} 
  \vcenter{ \def\w{20mm} \def\h{6mm} \xymatrix@C=0mm@R=0mm{
      {\scriptstyle q=1} & 
      *=<\w,\h>[F]{ 0 } & *=<\w,\h>[F]{ 0 } &
      *=<\w,\h>[F]{ \Rep{1} } 
      \\ {\scriptstyle q=0} & 
      *=<\w,\h>[F]{ \Rep{1} } & 
      *=<\w,\h>[F]{ 0 } & 
      *=<\w,\h>[F]{ 0 } 
      \\ & 
      {\scriptstyle p=0} & {\scriptstyle p=1} & {\scriptstyle p=2} 
    }} 
  \Rightarrow
  H^{p+q}\Big(\tx,\cO_\tx\Big)
  .
\end{equation}
From eqns.~\eqref{16} and~\eqref{19} we see that
\begin{equation} 
  H^3 \Big( \tx, \cO_\tx \Big) = \Hpq{\cO_\tx}{2}{1} = \Rep{1}
  ,
  \label{20}
\end{equation}
where the $\Rep{1}$ indicates that $H^3 ( \tx, \cO_{\tx} )$ is a
one-dimensional space carrying the trivial action of $\z3z3$.

\subsection{The First Leray Decomposition of Higgs Fields}

Now consider the $(p,q)$ Leray tableau for the sheaf $\cF = \atv$,
which is the second term in eq.~\eqref{14}. This was explicitly
computed in~\cite{ModHetSM} and is given by
\begin{equation}
  \label{22} 
  \vcenter{ \def\w{20mm} \def\h{6mm} \xymatrix@C=0mm@R=0mm{
      {\scriptstyle q=1} & 
      *=<\w,\h>[F]{ 0 } & 
      *=<\w,\h>[F]{ \rho_{14} } & 
      *=<\w,\h>[F]{ 0 } 
      \\ {\scriptstyle q=0} & 
      *=<\w,\h>[F]{ 0 } &
      *=<\w,\h>[F]{ \rho_{14} } & 
      *=<\w,\h>[F]{ 0 } \\ &
      {\scriptstyle p=0} & {\scriptstyle p=1} & {\scriptstyle p=2} 
    }} 
  \Rightarrow 
  H^{p+q}\Big(\Xt, \atv\Big)
  ,
\end{equation}
where $\rho_{14}$ is the fourteen-dimensional representation
\begin{equation} 
  \rho_{14}=
  \Big( 
  1 \oplus \chi_{1} \oplus
  \chi_{2} \oplus \chi_{1}^{2} \oplus \chi_{2}^{2} \oplus \chi_{1}
  \chi_{2}^{2} \oplus \chi_{1}^{2} \chi_{2} 
  \Big)^{\oplus 2}
  \label{23}
\end{equation}
of $\z3z3$. In general, it follows from eq.~\eqref{16} that $H^1(\tx,
\atv )$ is the sum of the two subspaces $\Hpq{\atv}{0}{1} \oplus
\Hpq{\atv}{1}{0}$.  However, we see from the Leray tableau
eq.~\eqref{22} that the $\Hpq{\atv}{0}{1}$ space vanishes. Hence,
\begin{equation} 
  H^1\Big(\tx, \atv \Big) = \Hpq{\atv}{1}{0}
  .
  \label{24}
\end{equation}
Furthermore, eq.~\eqref{22} tells us that
\begin{equation}
  \Hpq{\atv}{1}{0} = \rho_{14}
  .
  \label{25}
\end{equation}
%

\subsection{The First Leray Decomposition of the Moduli}

The (tangent space to the) moduli space of the vector bundle $\Vt$ is
$H^1(\tx, \ad(\tv))$, the first term in eq.~\eqref{14}. First, note
that $\ad(\tv)$ is defined to be the traceless part of $\tv \otimes
\tv^\dual$. But the trace is just the trivial line bundle, whose first
cohomology group vanishes. Therefore
\begin{equation}
  \label{eq:tensadtrace} 
  H^1\Big(\tx, \ad(\tv)\Big) = 
  H^1\Big(\tx, \Vt \otimes \Vt^\dual \Big) - 
  {\underbrace{H^1\Big(\tx, \cO_\tx \Big)}_{=0}} 
  .
\end{equation}
Since the action of the Wilson line on the ${\bf 1}$ representation of
$Spin(10)$ is trivial, one need only consider the $\z3z3$ invariant
subspaces of these cohomologies. That is, in the decomposition of the
index of the Dirac operator, eq.~\eqref{12}, the moduli fields are
contained in
\begin{equation}
  \left( H^1\big(\tx, \ad(\tv) \big) \otimes
    \Rep{1} \right)^{\z3z3}   
  = 
  H^1\Big(\tx, \ad(\tv) \Big)^{\z3z3}   
  =
  H^1\Big(\tx, \tv\otimes\tv^\dual \Big)^\z3z3
  .
\end{equation}
In a previous paper~\cite{ModHetSM}, we computed the total number of
moduli, but not their $(p,q)$ degrees.  However, this can be
calculated in a straightforward manner.

To compute $H^1(\tx, \Vt \otimes \Vt^\dual)^\z3z3$, recall
from~\cite{ModHetSM} that the short exact bundle sequence
eq.~\eqref{8} generates a complex of intertwined long exact sequences
which can be schematically represented by
\begin{equation}
  \label{26} 
  \vcenter{\xymatrix@R=5mm@C=8mm{
      *+[F--]{ H^*\Big(V_2 \otimes V_1^\dual\Big)^\z3z3 } \ar[d] \ar[r] & 
      H^*\Big(\Vt \otimes V_1^\dual\Big)^\z3z3 \ar[d] \ar[r] & 
      *+[F--]{ H^*\Big(V_1 \otimes V_1^\dual\Big)^\z3z3 } \ar[d] \\ 
      H^*\Big(V_2 \otimes \Vt^\dual\Big)^\z3z3 \ar[d] \ar[r] & 
      H^*\Big(\Vt \otimes \Vt^\dual\Big)^\z3z3 \ar[d] \ar[r] &
      H^*\Big(V_1 \otimes \Vt^\dual\Big)^\z3z3 \ar[d] \\ 
      *+[F--]{ H^*\Big(V_2 \otimes V_2^\dual\Big)^\z3z3 } \ar[r] & 
      H^*\Big(\Vt \otimes V_2^\dual\Big)^\z3z3 \ar[r] & 
      *+[F--]{ H^*\Big(V_1 \otimes V_2^\dual\Big)^\z3z3 
        ,        
      } 
    }} 
\end{equation} 
where $*$ means the complete cohomology with $*=0,1,2,3$ and we have
suppressed the base manifold $\tx$ for notational simplicity.  The
$(p,q)$ Leray tableaux for the ``corner'' cohomologies, marked by the
dashed boxes in eq.~\eqref{26}, were calculated in~\cite{ModHetSM}.
Actually, the whole cohomology groups were determined, not just their
invariant part.  Restricting to the $\z3z3$-invariant subspace, we
obtain
\begin{subequations}
  \begin{align}
    \label{27} 
    \vcenter{ \def\w{20mm} \def\h{6mm} \xymatrix@C=0mm@R=0mm{
        {\scriptstyle q=1} & *=<\w,\h>[F]{ \Rep{0} } & *=<\w,\h>[F]{
          \Rep{0} } & *=<\w,\h>[F]{ \Rep{4} } \\ {\scriptstyle q=0} &
        *=<\w,\h>[F]{ \Rep{4}} & *=<\w,\h>[F]{ \Rep{0} } & *=<\w,\h>[F]{
          \Rep{0} } \\ & {\scriptstyle p=0} & {\scriptstyle p=1} &
        {\scriptstyle p=2} }}
    & \, \Rightarrow H^{p+q}\Big(\Xt,V_1 \otimes V_1^\dual\Big)^\z3z3 ,
    \\
    \label{28} 
    \vcenter{ \def\w{20mm} \def\h{6mm} \xymatrix@C=0mm@R=0mm{
        {\scriptstyle q=1} & *=<\w,\h>[F]{ \Rep{4} } & *=<\w,\h>[F]{
          \Rep{16} } & *=<\w,\h>[F]{ \Rep{0} } \\ {\scriptstyle q=0} &
        *=<\w,\h>[F]{ \Rep{0} } & *=<\w,\h>[F]{ \Rep{0} } &
        *=<\w,\h>[F]{ \Rep{0} } \\ & {\scriptstyle p=0} & {\scriptstyle
          p=1} & {\scriptstyle p=2} }} 
    & \, \Rightarrow H^{p+q}\Big(\Xt,V_1 \otimes V_2^\dual\Big)^\z3z3 ,
    \\
    \label{29} 
    \vcenter{ \def\w{20mm} \def\h{6mm} \xymatrix@C=0mm@R=0mm{
        {\scriptstyle q=1} & *=<\w,\h>[F]{ \Rep{0} } & *=<\w,\h>[F]{
          \Rep{0} } & *=<\w,\h>[F]{ \Rep{0} } \\ {\scriptstyle q=0} &
        *=<\w,\h>[F]{ \Rep{0} } & *=<\w,\h>[F]{ \Rep{16} } &
        *=<\w,\h>[F]{ \Rep{4} } \\ & {\scriptstyle p=0} &
        {\scriptstyle p=1} & {\scriptstyle p=2} }} 
    & \, \Rightarrow H^{p+q}\Big(\Xt,V_2 \otimes V_1^\dual\Big)^\z3z3
    ,
    \\
    \label{30} 
    \vcenter{ \def\w{20mm} \def\h{6mm} \xymatrix@C=0mm@R=0mm{
        {\scriptstyle q=1} & *=<\w,\h>[F]{ \Rep{0} } & *=<\w,\h>[F]{
          \Rep{3} } & *=<\w,\h>[F]{ \Rep{1} } \\ {\scriptstyle q=0} &
        *=<\w,\h>[F]{ \Rep{1} } & *=<\w,\h>[F]{ \Rep{3} } &
        *=<\w,\h>[F]{ \Rep{0} } \\ & {\scriptstyle p=0} & {\scriptstyle
          p=1} & {\scriptstyle p=2} }} 
    & \, \Rightarrow H^{p+q}\Big(\Xt,V_2 \otimes
    V_2^\dual\Big)^\z3z3 
    ,
  \end{align}
\end{subequations}
where, as above, the $\Rep{3}$, $\Rep{4}$, and $\Rep{16}$ denote the
rank $3$, $4$, and $16$ trivial representation of $\z3z3$.
Furthermore, the $H^0$ and, by Serre duality, the $H^3$ entries in the
$(p,q)$ Leray tableaux for the remaining cohomology groups in
eq.~\eqref{26}
were computed in~\cite{ModHetSM}, where it was found that
\begin{subequations}
\begin{align}
  \label{31} 
  \vcenter{ \def\w{20mm} \def\h{6mm} \xymatrix@C=0mm@R=0mm{
      {\scriptstyle q=1} & *=<\w,\h>[F]{ ** } & *=<\w,\h>[F]{ ** } &
      *=<\w,\h>[F]{ \Rep{4} } \\ {\scriptstyle q=0} & *=<\w,\h>[F]{ \Rep{0} }
      & *=<\w,\h>[F]{ ** } & *=<\w,\h>[F]{ ** } \\ & {\scriptstyle p=0} &
      {\scriptstyle p=1} & {\scriptstyle p=2} }}
  & \,  \Rightarrow  H^{p+q}\Big(\Xt,\Vt \otimes V_1^\dual\Big)^\z3z3  ,
  \\
  \label{32} 
  \vcenter{ \def\w{20mm} \def\h{6mm} \xymatrix@C=0mm@R=0mm{
      {\scriptstyle q=1} & *=<\w,\h>[F]{ ** } & *=<\w,\h>[F]{ ** } &
      *=<\w,\h>[F]{ \Rep{0} } \\ {\scriptstyle q=0} & *=<\w,\h>[F]{
        \Rep{4} } & *=<\w,\h>[F]{ ** } & *=<\w,\h>[F]{ ** } \\ &
      {\scriptstyle p=0} & {\scriptstyle p=1} & {\scriptstyle p=2} }}
  & \, \Rightarrow H^{p+q}\Big(\Xt,V_1 \otimes \Vt^\dual\Big)^\z3z3 ,
  \\
  \label{33} 
  \vcenter{ \def\w{20mm} \def\h{6mm} \xymatrix@C=0mm@R=0mm{
      {\scriptstyle q=1} & *=<\w,\h>[F]{ ** } & *=<\w,\h>[F]{ ** } &
      *=<\w,\h>[F]{ \Rep{0} } \\ {\scriptstyle q=0} & *=<\w,\h>[F]{ \Rep{1} }
      & *=<\w,\h>[F]{ ** } & *=<\w,\h>[F]{ ** } \\ & {\scriptstyle p=0} &
      {\scriptstyle p=1} & {\scriptstyle p=2} }} 
  & \,  \Rightarrow  H^{p+q}\Big(\Xt,\Vt \otimes V_2^\dual\Big)^\z3z3  ,
  \\
  \label{34} 
  \vcenter{ \def\w{20mm} \def\h{6mm} \xymatrix@C=0mm@R=0mm{
      {\scriptstyle q=1} & *=<\w,\h>[F]{ ** } & *=<\w,\h>[F]{ ** } &
      *=<\w,\h>[F]{ \Rep{1} } \\ {\scriptstyle q=0} & *=<\w,\h>[F]{ \Rep{0} }
      & *=<\w,\h>[F]{ ** } & *=<\w,\h>[F]{ ** } \\ & {\scriptstyle p=0} &
      {\scriptstyle p=1} & {\scriptstyle p=2} }} 
  & \,  \Rightarrow  H^{p+q}\Big(\Xt,V_2 \otimes\Vt^\dual\Big)^\z3z3  ,
  \\
  \label{35} 
  \vcenter{ \def\w{20mm} \def\h{6mm} \xymatrix@C=0mm@R=0mm{
      {\scriptstyle q=1} & *=<\w,\h>[F]{ ** } & *=<\w,\h>[F]{ ** } &
      *=<\w,\h>[F]{ \Rep{1} } \\ {\scriptstyle q=0} & *=<\w,\h>[F]{
        \Rep{1} } & *=<\w,\h>[F]{ ** } & *=<\w,\h>[F]{ ** } \\ &
      {\scriptstyle p=0} & {\scriptstyle p=1} & {\scriptstyle p=2} }}
  & \, \Rightarrow H^{p+q}\Big(\Xt,\Vt \otimes \Vt^\dual\Big)^\z3z3 .
\end{align}
\end{subequations}
The cohomology spaces on $B_2$ which are thus far uncalculated are
denoted by $**$.

Our goal is to compute the entries in the $(p,q)$ Leray tableaux for
$H^1(\tv \otimes \tv^{\dual})^{\z3z3}$ at the positions $(0,1)$ and
$(1,0)$ in eq.~\eqref{35}. This can be accomplished as follows. First
consider the $\z3z3$ invariant part of the lower horizontal long exact
sequence in eq.~\eqref{26}.  Restricting ourselves to the entries
contributing to $H^1$, the exact sequence reads
\begin{equation}
  \label{eq:moduliH1les}
  \vcenter{\xymatrix@R=10pt@M=4pt@H+=22pt@C=8mm{
      & 
      &
      \cdots
      \ar[r] &
      H^0\big(V_1 \otimes V_2^\dual\big)^\z3z3
      \ar`[rd]`[l]`[dlll]`[d][dll] & 
      \\
      & 
      H^1\big(V_2 \otimes V_2^\dual\big)^\z3z3
      \ar[r] &
      H^1\big(\Vt \otimes V_2^\dual\big)^\z3z3
      \ar[r] &
      H^1\big(V_1 \otimes V_2^\dual\big)^\z3z3
      \ar`[rd]`[l]`[dlll]^(0.2){\delta_1^\dual}`[d][dll] & 
      \\
      & 
      H^2\big(V_2 \otimes V_2^\dual\big)^\z3z3
      \ar[r] &
      \cdots
      .
      &
      &
    }}
\end{equation}
In~\cite{ModHetSM} it was proven that 
\begin{equation}
  \label{badd1}
  H^0\big(V_1 \otimes V_2^\dual\big)^\z3z3 =0
  ,\qquad
  \delta_{1}^{\dual}=0
  .  
\end{equation}
Hence, both coboundary maps vanish and we obtain the short exact
sequence
\begin{equation} 
  \label{36}
  \vcenter{\xymatrix@C=5mm@R=5mm{ 
      0 \ar@{=}[d]\ar[r] & 
      H^1(V_2 \otimes V_2^\dual)^\z3z3  \ar@{<=}[d]\ar[r] & 
      H^1(\Vt \otimes V_2^\dual)^\z3z3  \ar@{<=}[d]\ar[r] & 
      H^1(V_1 \otimes V_2^\dual)^\z3z3  \ar@{<=}[d]\ar[r] & 
      0 \ar@{=}[d]
      \\
      0 \ar[r] &
      {\begin{array}{|c|c|c|}\hline 
          \Rep{0} &~&~ \\ \hline ~& \Rep{3} &~ \\ \hline 
        \end{array}} \ar[r]_{\phi_1} & 
      {\begin{array}{|c|c|c|}\hline
          ** &~&~ \\ \hline ~& ** &~ \\ \hline 
        \end{array}} \ar[r]_{\phi_2} &
      {\begin{array}{|c|c|c|}\hline 
          \Rep{4} &~&~ \\ \hline ~& \Rep{0} &~ \\ \hline 
        \end{array}} \ar[r] &
      0
      .
    }} 
\end{equation}
%
%
Now, on general grounds the coboundary maps in a long exact sequence
increase the cohomology degree, while the interior maps preserve the
cohomology degree. In particular, the maps $\phi_1$ and $\phi_2$ in
eq.~\eqref{36} must preserve the $(p,q)$ degrees. The $(0,1)$ and
$(1,0)$ entries in the $H^*(\tv \otimes V_{2}^{\dual})^{\z3z3}$ Leray
tableau can now be evaluated using the following general formula.
Consider an exact sequence of linear spaces
\begin{equation} 
  \dots \longrightarrow {\cal{U}}
  \stackrel{m_{1}}{\longrightarrow} {\cal{V}} \longrightarrow {\cal{W}}
  \longrightarrow {\cal{X}} \stackrel{m_{2}}{\longrightarrow} {\cal{Y}}
  \longrightarrow \dots \,,
  \label{37}
\end{equation}
where $m_{1}$ and $m_{2}$ are coboundary maps.  Then
\begin{equation} 
  \dim_\C({\cal{W}})= \dim_\C({\cal{V}}) +
  \dim_\C({\cal{X})} - \rank (m_{1}) -\rank (m_{2}).
  \label{38}
\end{equation}
This formula applies to any linear spaces, such as entire cohomology
groups or their individual $(p,q)$ Leray subspaces. Using
eq.~\eqref{38} for the $(0,1)$ and $(1,0)$ Leray degrees separately in
eq.~\eqref{36}, we obtain the desired entries in the Leray tableau
\begin{equation}
  \label{39} 
  \vcenter{ \def\w{20mm} \def\h{6mm} \xymatrix@C=0mm@R=0mm{
      {\scriptstyle q=1} & *=<\w,\h>[F]{ \Rep{4} } & *=<\w,\h>[F]{ } &
      *=<\w,\h>[F]{ } \\ {\scriptstyle q=0} & *=<\w,\h>[F]{ } &
      *=<\w,\h>[F]{ \Rep{3} } & *=<\w,\h>[F]{ } \\ & {\scriptstyle p=0}
      & {\scriptstyle p=1} & {\scriptstyle p=2} }} 
  \Rightarrow 
  H^{p+q}\Big(\Xt, \tv \otimes V_{2}^{\dual}\Big)^{\z3z3}
  .
\end{equation}
Second, consider the upper horizontal long exact sequence in
eq.~\eqref{26}. Restricting ourselves to the entries contributing to
$H^1$, this is given by
\begin{equation} 
  \vcenter{\xymatrix@C=10pt{ 
    {\scriptstyle H^0(V_1 \otimes V_1^\dual)^\z3z3 } \ar@{<=}[d]\ar[rr]^{d_2} &&
    {\scriptstyle H^1(V_2 \otimes V_1^\dual)^\z3z3 } \ar@{<=}[d]\ar[r] & 
    {\scriptstyle H^1(\Vt \otimes V_1^\dual)^\z3z3 } \ar@{<=}[d]\ar[r] & 
    {\scriptstyle H^1(V_1 \otimes V_1^\dual)^\z3z3 } \ar@{<=}[d]\ar[r] & 
    \cdots
    \\
    {\begin{array}{|c|c|c|}\hline ~&~&~ \\
        \hline \Rep{4} &~&~ \\ \hline \end{array}} 
    \ar@<+8pt>[rr]^<>(0.37){d_2|_{(0,1)}} 
    \ar@<-8pt>[rr]_<>(0.37){d_2|_{(1,0)}} 
    &&
    {\begin{array}{|c|c|c|}\hline \Rep{0} &~&~ \\ \hline ~&\Rep{16}&~ \\
        \hline \end{array}} \ar[r] & 
    {\begin{array}{|c|c|c|}\hline ** &~&~ \\
        \hline ~& ** &~ \\ \hline \end{array}} \ar[r] &
    {\begin{array}{|c|c|c|}\hline \Rep{0} &~&~ \\ \hline ~& \Rep{0} &~ \\
        \hline \end{array}} \ar[r] & \cdots 
    .
    \\ 
  }} 
  \label{40}
\end{equation}
The coboundary map $d_{2}$ on the left was shown in~\cite{ModHetSM} to
have $\rank(d_{2})=4$. In the context of the $(p,q)$ Leray tableaux,
it decomposes as
\begin{equation} 
  \label{41}
  \rank\Big( d_2|_{(0,1)}: \Rep{4} \to \Rep{0} \Big) = 0
  , \quad 
  \rank\Big( d_2|_{(1,0)}: \Rep{4} \to \Rep{16} \Big) = 4
  .
\end{equation}
Again using eq.~\eqref{38} for the $(0,1)$ and $(1,0)$ Leray degrees
separately in eq.~\eqref{40}, we obtain the desired entries in the
Leray tableau
\begin{equation}
  \label{42} 
  \vcenter{ \def\w{20mm} \def\h{6mm} \xymatrix@C=0mm@R=0mm{
      {\scriptstyle q=1} & *=<\w,\h>[F]{ {\bf 0} } & *=<\w,\h>[F]{ } &
      *=<\w,\h>[F]{ } \\ {\scriptstyle q=0} & *=<\w,\h>[F]{ } &
      *=<\w,\h>[F]{ {\bf 12} } & *=<\w,\h>[F]{ } \\ & {\scriptstyle p=0}
      & {\scriptstyle p=1} & {\scriptstyle p=2} }} 
  \Rightarrow 
  H^{p+q}\Big(\Xt, \tv \otimes V_{1}^{\dual}\Big)^{\z3z3}
  .
\end{equation}
From the results in eqns.~\eqref{39} and~\eqref{42}, we can finally
compute the $(p,q)$ Leray subspaces that determine $H^1(\tv
\otimes \tv^{\dual})^{\z3z3}$ in eq.~\eqref{35} using the middle
vertical exact sequence of eq.~\eqref{26}
\begin{equation} 
  \vcenter{\xymatrix@C=20pt@R=10pt{ 
    \cdots \ar[r]^-{d_3} & 
    H^1(\Vt \otimes V_1^\dual)^\z3z3  \ar@{<=}[d]\ar[r] & 
    H^1(\Vt \otimes \Vt^\dual)^\z3z3  \ar@{<=}[d]\ar[r] & 
    H^1(\Vt \otimes V_2^\dual)^\z3z3  \ar@{<=}[d]\ar[r]^-{\delta_2} & 
    \cdots
    \\
    \cdots \ar[r] & 
    {\begin{array}{|c|c|c|}\hline \Rep{0} &~&~ \\ \hline ~&\Rep{12} &~ \\
        \hline \end{array}} \ar[r] & 
    {\begin{array}{|c|c|c|}\hline ** &~&~ \\
        \hline ~& ** &~ \\ \hline \end{array}} \ar[r] &
    {\begin{array}{|c|c|c|}\hline \Rep{4} &~&~ \\ \hline ~&\Rep{3}&~ \\
        \hline \end{array}} \ar[r] & 
    \cdots
    .
  }}
  \label{43}
\end{equation}
In~\cite{ModHetSM}, we calculated both coboundary maps $d_3$ and
$\delta_2$. It was found that they both vanish, that is
\begin{equation} 
  d_3 =
  \delta_2 = 0
  .
  \label{badd2}
\end{equation}
Using these results and eq.~\eqref{38} for each of the two $H^1$ Leray
subspace sequences in eq.~\eqref{43}, we find that the $H^1$ entries
in the Leray tableau for $H^{*}(\Xt, \tv \otimes \tv^{\dual})^{\z3z3}$ are
\begin{equation}
  \label{44} \vcenter{ \def\w{20mm} \def\h{6mm} \xymatrix@C=0mm@R=0mm{
      {\scriptstyle q=1} & *=<\w,\h>[F]{ {\bf 4} } & *=<\w,\h>[F]{ } &
      *=<\w,\h>[F]{ } \\ {\scriptstyle q=0} & *=<\w,\h>[F]{ } &
      *=<\w,\h>[F]{ {\bf 15} } & *=<\w,\h>[F]{ } \\ & {\scriptstyle p=0}
      & {\scriptstyle p=1} & {\scriptstyle p=2} }} 
  \Rightarrow
  H^{p+q}\Big(\Xt, \tv \otimes \tv^{\dual}\Big)^{\z3z3}
  .
\end{equation}
Note that
\begin{equation} 
  h^1(\Xt, \tv \otimes \tv^{\dual})^{\z3z3} =4+15=19,
  \label{45}
\end{equation}
which is consistent with the conclusion in~\cite{ModHetSM} that there
are a total of nineteen vector bundle moduli. Now, however, we have
determined the $(p,q)$ decomposition of $H^1(\Xt, \tv \otimes
\tv^{\dual})^{\z3z3}$ into the subspaces
\begin{equation} 
  \label{46}
  H^1\Big(\Xt,\tv \otimes \tv^{\dual}\Big)^{\z3z3}=
  \Hpq{\tv \otimes \tv^{\dual}}{0}{1}^{\z3z3}
  ~\oplus~
  \Hpq{\tv \otimes \tv^{\dual}}{1}{0}^{\z3z3}
  ,
\end{equation}
where
\begin{equation} 
  \Hpq{\tv \otimes \tv^{\dual}}{0}{1}^{\z3z3} = \Rep{4}
  ,\quad
  \Hpq{\tv \otimes \tv^{\dual}}{1}{0}^{\z3z3} = \Rep{15}  
  \label{47}
\end{equation}
respectively.

\subsection{The (p,q) Selection Rule}

Having computed the decompositions of $H^3(\tx, \cO_{\tx} )$,
$H^1(\tx, \atv )$ and $H^1(\tx, \ad(\tv) )^{\z3z3}$ into their
${(p,q)}$ Leray subspaces, we can now analyze the $(p,q)$ components
of the triple product 
\begin{equation}
  \label{eq:product}
  H^1\Big(\tx, \Vt \otimes \Vt^\dual \Big)^\z3z3 \otimes 
  H^1\Big(\tx, \atv \Big) \otimes
  H^1\Big(\tx, \atv \Big) 
  \longrightarrow 
  H^3\Big(\tx, \cO_{\tx} \Big)  
\end{equation}
given in eq.~\eqref{eq:cup}. Inserting eqns.~\eqref{24}
and~\eqref{46}, we see that
\begin{multline}
  \label{eq:pqCup}
  H^1\Big(\tx, \Vt \otimes \Vt^\dual \Big)^\z3z3 \otimes 
  H^1\Big(\tx, \atv \Big) \otimes
  H^1\Big(\tx, \atv \Big) 
  = \\ =
  \Big( \Hpq{\Vt \otimes \Vt^\dual}{0}{1} \oplus 
        \Hpq{\Vt \otimes \Vt^\dual}{1}{0} \Big)
  \otimes
  \Hpq{\atv}{1}{0} \otimes 
  \Hpq{\atv}{1}{0} 
  = \\ =
  \underbrace{\Big( 
    {\scriptstyle 
      \Hpq{\Vt \otimes \Vt^\dual}{0}{1}^\z3z3 \otimes
      \Hpq{\atv}{1}{0} \otimes 
      \Hpq{\atv}{1}{0} 
    }
    \Big) 
  }_{\text{total $(p,q)$ degree }=\, (2,1)}
  \oplus
  \underbrace{\Big( 
    {\scriptstyle 
      \Hpq{\Vt \otimes \Vt^\dual}{1}{0}^\z3z3 \otimes
      \Hpq{\atv}{1}{0} \otimes 
      \Hpq{\atv}{1}{0} 
    }
    \Big) 
  }_{\text{total $(p,q)$ degree }=\, (3,0)}
  .
\end{multline}
Because of the $(p,q)$ degree, only the first term can have a
non-zero product in
\begin{equation}
  \label{48}
  H^3 \Big( \tx, \cO_\tx \Big) = \Hpq{\cO_\tx}{2}{1}
  ,
\end{equation}
see eq.~\eqref{20}. It follows that out of the $H^1(\tv \otimes
\tv^{\dual})^{\z3z3}=\Rep{19}$ vector bundle moduli, only
\begin{equation} 
  \Hpq{\tv \otimes \tv^{\dual}}{0}{1}^{\z3z3} = \Rep{4}
  \label{49}
\end{equation}
will form non-vanishing Higgs--Higgs conjugate $\mu$-terms.  The
remaining fifteen moduli in the $\Hpq{\tv \otimes
  \tv^{\dual}}{1}{0}^{\z3z3}$ component have the wrong $(p,q)$ degree
to couple to a Higgs--Higgs conjugate pair.  We refer to this as the
\emph{$(p,q)$ Leray degree selection rule}. We conclude that the only
non-zero product in eq.~\eqref{eq:product} is of the form
\begin{equation}
  \label{eq:pqproduct}
  \Hpq{\Vt \otimes \Vt^\dual}{0}{1}^\z3z3 \otimes
  \Hpq{\atv}{1}{0} \otimes 
  \Hpq{\atv}{1}{0} 
  \longrightarrow
  \Hpq{\cO_\Xt}{2}{1}
  .
\end{equation}
Roughly what happens is the following. The Leray spectral sequence
decomposes differential forms into the number $p$ of legs in the
direction of the base and the number $q$ of legs in the fiber
direction. The holomorphic $(3,0)$-form $\Omega$ has two legs in the
base and one leg in the fiber direction. According to eq.~\eqref{24},
both $1$-forms $\Psi_H$, $\Psi_{\bar{H}}$ corresponding to Higgs and
Higgs conjugate have their one leg in the base direction.  Therefore,
the wedge product in eq.~\eqref{eq:lambdaintegral} can only be non-zero if the
modulus $1$-form $\Psi_\phi$ has its leg in the fiber direction, which
only $4$ out of the $19$ moduli satisfy.

We conclude that due to a selection rule for the $(p,q)$ Leray degree,
the Higgs $\mu$-terms in the effective low energy theory can involve
only four of the nineteen vector bundle moduli.

\section{The Second Elliptic Fibration}
\label{sec:LSS2}

So far, we only made use of the fact that our Calabi-Yau manifold is
an elliptic fibration over the base $B_2$. But the $\dP9$ surface
$B_2$ is itself elliptically fibered over a $\IP^1$. Consequently,
there is yet another selection rule coming from the second elliptic
fibration. 

Therefore, we now consider the second Leray spectral sequence
corresponding to the projection
\begin{equation} 
  B_2 \stackrel{\beta_{2}}{\longrightarrow}
  \IP^1.
  \label{50}
\end{equation}
For any sheaf $\cFt$ on $B_2$, the Leray sequence tells
us that
\begin{equation} 
  H^p\Big(B_2, \cFt \Big)=
  \bigoplus^{s+t=p}_{s,t} H^{s} \Big( \IP^1, R^{t} \beta_{2*}
  \cFt \Big),
  \label{51}
\end{equation}
where the only non-vanishing entries are for $s=0,1$ (since $\dim_\C
\IP^1=1$) and $t=0,1$ (since the fiber of $B_2$ is an elliptic curve).
The cohomologies $H^{s} ( \IP^1, R^{t} \beta_{2*} \cFt)$ fill out the
$2 \times 2$ Leray tableau
\begin{equation}
  \label{52} 
  \vcenter{ \def\w{40mm} \def\h{8mm} \xymatrix@C=0mm@R=0mm{
      {\scriptstyle t=1} & 
      *=<\w,\h>[F]{ H^0( \IP^1, R^1\beta_{2\ast} \cFt ) } & 
      *=<\w,\h>[F]{ H^1( \IP^1, R^1\beta_{2\ast}\cFt) } \\ 
      {\scriptstyle t=0} &
      *=<\w,\h>[F]{ H^0( \IP^1, \beta_{2\ast} \cFt ) } &
      *=<\w,\h>[F]{ H^1( \IP^1, \beta_{2\ast} \cFt ) } \\ &
      {\scriptstyle s=0} & {\scriptstyle s=1} }} 
  \Rightarrow
  H^{s+t}\Big(B_2, \cFt \Big)
  .
\end{equation}
As is clear from eq.~\eqref{51}, the sum over the diagonals yields the
desired cohomology of $\cFt$.  Note that to evaluate the product
eq.~\eqref{eq:pqproduct}, we need the $[s,t]$ Leray tableaux for
\begin{equation}
  \cFt = 
  R^1\pi_{2\ast} \big(\Vt \otimes \Vt^\dual \big)
  ,\quad
  \pi_{2\ast} \big( \atv \big)
  ,\quad
  R^1\pi_{2\ast} \big(\cO_\Xt\big)
  .
\end{equation}
In the following, it will be useful to define
\begin{equation} 
  H^{s} \bigg( \IP^1, 
  R^t \beta_{2*} \Big( R^q \pi_{2\ast}\big(\cF\big)  \Big) \bigg) 
  \equiv 
  \Hst{\cF}{q}{s}{t}
  .
  \label{53}
\end{equation}
One can think of $\Hst{\cF}{q}{s}{t}$ as the subspace of
$H^\ast\big(\Xt, \cF\big)$ that can be written as forms with $q$ legs
in the $\pi_2$-fiber direction, $t$ legs in the $\beta_2$-fiber
direction, and $s$ legs in the base $\IP^1$ direction.


\subsection{The Second Leray Decomposition of the Volume Form}

Let us first discuss the $[s,t]$ Leray tableau for the sheaf
$\cFt=R^1\pi_{2\ast} \big(\cO_{\tx}\big)$. Since $R^1\pi_{2\ast}
\big(\cO_{\tx}\big)=K_{B_2}$, the canonical line bundle of $B_2$, it follows
immediately that
\begin{equation}
  \label{54} 
  \vcenter{ \def\w{20mm} \def\h{6mm} \xymatrix@C=0mm@R=0mm{
      {\scriptstyle t=1} & *=<\w,\h>[F]{ 0 } & *=<\w,\h>[F]{ \Rep{1} }
      \\ {\scriptstyle t=0} & *=<\w,\h>[F]{ 0 } & *=<\w,\h>[F]{ 0 } \\ &
      {\scriptstyle s=0} & {\scriptstyle s=1} }} 
  \Rightarrow
  H^{s+t}\Big(B_2, R^1\pi_{2\ast} \big(\cO_{\tx}\big) \Big)  
  .
\end{equation}
In our notation, this means that
\begin{equation}
  H^2\Big(B_2, R^1\pi_{2\ast} \big(\cO_{\tx}\big)\Big) = 
  \Hst{\cO_\Xt}{1}{1}{1}
\end{equation}
has pure $[s,t]=[1,1]$ degree. We see from eqns.~\eqref{54}
and~\eqref{51} that
\begin{equation} 
  \label{55}
  H^3\Big( \Xt, \cO_\Xt \Big) = 
  \Hpq{\cO_\Xt}{2}{1} =
  \Hst{\cO_\Xt}{1}{1}{1} =
  \Rep{1}
  .
\end{equation}

\subsection{The Second Leray Decomposition of Higgs Fields}

Now consider the $[s,t]$ Leray tableau for the sheaf
$\cFt=\pi_{2\ast}\big(\atv\big)$.  This was explicitly computed
in~\cite{ModHetSM} and is given by
\begin{equation}
  \label{57} 
  \vcenter{ \def\w{65mm} \def\h{8mm} \xymatrix@C=0mm@R=0mm{
      {\scriptstyle t=1} & 
      *=<\w,\h>[F]{ 
        (1 \oplus \chi_{1} \oplus \chi_{1}^{2} 
        \oplus \chi_{1}^{2} \oplus \chi_{2}^{2} \oplus
        \chi_{1}\chi_{2}^{2})^{\oplus2} 
      } & 
      *=<20mm,\h>[F]{ 0 } \\
      {\scriptstyle t=0} & 
      *=<\w,\h>[F]{ 0 } & 
      *=<20mm,\h>[F]{ 
        (\chi_{1}^{2}\chi_{2})^{\oplus2}}  
      \\ & {\scriptstyle s=0} &
      {\scriptstyle s=1} }} 
  \Rightarrow 
  H^{s+t}\Big(B_2, \pi_{2\ast} \big(\atv\big) \Big)    
  .
\end{equation}
This means that the $14$ copies of the $\Rep{10}$ of $Spin(10)$ given
in eq.~\eqref{25} split as
\begin{equation}
  H^1\Big( \Xt, \atv \Big) = 
  \Hpq{\atv}{1}{0} = 
  \Hst{\atv}{0}{0}{1} \oplus \Hst{\atv}{0}{1}{0}
  ,
\end{equation}
where
\begin{equation}
  \label{58}
  \begin{split}
    \Hst{\atv}{0}{0}{1} ~&=
    \big( 1 \oplus \chi_{1} \oplus \chi_{1}^{2} \oplus \chi_{1}^{2} 
    \oplus \chi_{2}^{2} \oplus \chi_{1}\chi_{2}^{2} \big)^{\oplus2}
    \\
    \Hst{\atv}{0}{1}{0} ~&=    
    \big( \chi_{1}^{2}\chi_{2} \big)^{\oplus2}
    .
  \end{split}
\end{equation}
Note that
\begin{equation} 
  \label{60}
  \Hst{\atv}{0}{0}{1} \oplus \Hst{\atv}{0}{1}{0} = \rho_{14}
\end{equation}
in eq.~\eqref{23}, as it must.

\subsection{The Second Leray Decomposition of the Moduli}

Finally, let us consider the $[s,t]$ Leray tableau for the moduli. We
have already seen that, due to the $(p,q)$ selection rule, only
\begin{equation}
  \Hpq{\tv \otimes \tv^{\dual}}{0}{1}^{\z3z3} = 
  \Rep{4}
  \quad \subset
  H^1\Big( \Xt, \Vt \otimes \Vt^\dual \Big)^\z3z3 
\end{equation}
out of the $19$ moduli can occur in the Higgs--Higgs conjugate
$\mu$-term. Therefore, we are only interested in the $[s,t]$
decomposition of this subspace, that is, the degree $0$ cohomology of
the sheaf $R^1\pi_{2\ast} \big( \Vt \otimes \Vt^\dual \big)$. The
corresponding Leray tableau is given by
\begin{equation}
  \label{61} 
  \vcenter{ \def\w{20mm} \def\h{6mm} \xymatrix@C=0mm@R=0mm{
      {\scriptstyle t=1} & 
      *=<\w,\h>[F]{  } & 
      *=<\w,\h>[F]{  } \\
      {\scriptstyle t=0} & 
      *=<\w,\h>[F]{ \Rep{4} } & 
      *=<\w,\h>[F]{  }
      \\ & {\scriptstyle s=0} & {\scriptstyle s=1} }} 
  \Rightarrow
  H^{s+t}\Big(B_2, R^1\pi_{2\ast} \big(\Vt\otimes\Vt^\dual\big)
  \Big)^\z3z3     
  ,
\end{equation}
where the empty boxes are of no interest for our purposes. It follows
that the $4$ moduli of interest have $[s,t]$ degree $[0,0]$,
\begin{equation} 
  \label{62}
  \Hpq{\tv \otimes \tv^{\dual}}{0}{1}^{\z3z3} = 
  \Hst{\tv \otimes \tv^{\dual}}{1}{0}{0}^{\z3z3} = 
  \Rep{4}
  .
\end{equation}

\subsection{The [s,t] Selection Rule}

Having computed the decompositions of the relevant cohomology spaces
into their $[s,t]$ Leray subspaces, we can now calculate the triple
product eq.~\eqref{eq:cup}. The $(p,q)$ selection rule dictates that
the only non-zero product is of the form eq.~\eqref{eq:pqproduct}. Now
split each term in this product into its $[s,t]$ subspaces, as given
in eqns.~\eqref{55},~\eqref{58}, and~\eqref{62} respectively. The
result is
\begin{multline} 
  \label{65}
  \Hst{\tv \otimes \tv^{\dual}}{1}{0}{0}^{\z3z3} 
  \otimes
  \Big( \Hst{\atv}{0}{0}{1} \oplus \Hst{\atv}{0}{1}{0} \Big)
  \otimes \\ \otimes
  \Big( \Hst{\atv}{0}{0}{1} \oplus \Hst{\atv}{0}{1}{0} \Big)
  \longrightarrow
  \Hst{\cO_\Xt}{1}{1}{1} 
  .
\end{multline}
Clearly, this triple product vanishes by degree unless we choose the
$\Hst{\atv}{0}{0}{1}$ from one of the $\Hpq{\atv}{1}{0}$ subspaces and
$\Hst{\atv}{0}{1}{0}$ from the other.  In this case, eq.~\eqref{65}
becomes
\begin{equation} 
  \label{65-1}
  \Hst{\tv \otimes \tv^{\dual}}{1}{0}{0}^{\z3z3} 
  \otimes
  \Hst{\atv}{0}{1}{0}
  \otimes
  \Hst{\atv}{0}{0}{1}
  \longrightarrow
  \Hst{\cO_\Xt}{1}{1}{1} 
  ,
\end{equation}
which is consistent.

\subsection{Wilson Lines}

Recall that we have, in addition to the $SU(4)$ instanton, also a
Wilson line\footnote{In fact, we switch on a separate Wilson line for
  both $\Z_3$ factors in $\pi_1(X)=\z3z3$.} turned on. Its effect is
to break the $Spin(10)$ gauge group down to the desired $SU(3)_{C}
\times SU(2)_{L} \times U(1)_{Y} \times U(1)_{B-L}$ gauge group. Each
fundamental matter field in the $\Rep{10}$ can be broken to a Higgs
field, a color triplet, or projected out. In particular, we are going
to choose the Wilson line $W$ so that its $\z3z3$ action on a
$\Rep{10}$ representation of $Spin(10)$ is given by
\begin{equation} 
  \Rep{10}=
  \Big( 
  \chi_{1} \chi_{2}^{2} H \oplus \chi_{1} \chi_{2} C
  \Big) \oplus \Big( 
  \chi_{1}^{2} \chi_{2} \bar{H} \oplus \chi_{1}^{2} \chi_{2}^{2}
  \bar{C}
  \Big)
  ,
  \label{66}
\end{equation}
where
\begin{equation} 
  H=\big( \Rep{1},\Rep{2}, 3, 0 \big), \quad 
  C=\big( \Rep{3},\Rep{1}, -2, -2 \big)
  \label{67}
\end{equation}
are the Higgs and color triplet representations of $SU(3)_{C} \times
SU(2)_{L} \times U(1)_{Y} \times U(1)_{B-L}$
respectively.\footnote{The attentive reader will note that the $\z3z3$
  action of the Wilson line presented here differs from that given
  in~\cite{ModHetSM}.  Be that as it may, the low energy spectra of
  the two different actions are identical.  However, for the $\z3z3$
  action presented in this paper, there are non-vanishing Higgs
  $\mu$-terms whereas all $\mu$-terms vanish identically using the
  Wilson line action given in~\cite{ModHetSM}.} Tensoring this with
the cohomology space $H^1\big(\Xt,\atv\big)$, we find the invariant
subspace under the combined $\z3z3$ action on the cohomology space and
the Wilson line to be
\begin{equation}
  \Big[ H^1\big(\Xt,\atv\big) \otimes \Rep{10} \Big]^\z3z3
  = 
  \Span
  \big\{ H_1, H_2, \bar{H}_1, \bar{H}_2 \big\}
  .
\end{equation}
Hence, we find precisely two copies of Higgs and two copies of Higgs
conjugate fields survive the $\z3z3$ quotient. As required for any
realistic model, all color triplets are projected out.

The new information now are the $(p,q)$ and $[s,t]$ degrees of the
Higgs fields. Using the decomposition of $H^1\big(\Xt,\atv\big)$, we
find
\begin{multline}
  \Big[ H^1\big(\Xt,\atv\big) \otimes \Rep{10} \Big]^\z3z3
  =
  \Big[ \Hpq{\atv}{1}{0} \otimes \Rep{10} \Big]^\z3z3
  = \\ =
  \underbrace{
    \Big[ \Hst{\atv}{0}{0}{1} \otimes \Rep{10} \Big]^\z3z3
  }_{= \Span\{\bar{H}_1, \bar{H}_2\} }
  \oplus 
  \underbrace{
    \Big[ \Hst{\atv}{0}{1}{0} \otimes \Rep{10} \Big]^\z3z3
  }_{= \Span\{H_1, H_2\} }
  .
\end{multline}
The resulting degrees under the two Leray spectral sequences of the
Higgs and Higgs conjugate fields are listed in Table~\ref{tab:Higgsdegree}.
\begin{table}[htbp]
  \centering
  \begin{tabular}{c|cc}
    Field & $(p,q)$ & $[s,t]$
    \\ \hline
    $H_1, H_2$ & 
    $(1,0)$ & 
    $[1,0]$ 
    \\
    $\bar{H}_1, \bar{H}_2$ & 
    $(1,0)$ & 
    $[0,1]$ 
  \end{tabular}
  \caption{Degrees of the Higgs fields.}
  \label{tab:Higgsdegree}
\end{table}

\section{\texorpdfstring{Higgs $\mu$-terms}{Higgs mu-terms}}

To conclude, we analyzed cubic terms in the superpotential of the form
\begin{equation} 
  \lambda_{iab} \phi_{i} H_{a} \bar{H}_{b}
  ,
  \label{67-1}
\end{equation}
where
\begin{itemize}
\item $\lambda_{iab}$ is a coefficient determined by the integral
  eq.~\eqref{eq:lambdaintegral},
\item $\phi_i,~ i=1,\dots,19$ are the vector bundle moduli,
\item $H_a,~ a=1,2$ are the two Higgs fields, and
\item $\bar{H}_b,~  b=1,2$ are the two Higgs conjugate fields.
\end{itemize}
We found that they are subject to two independent selection rules
coming from the two independent torus fibrations. The first selection
rule is that the total $(p,q)$ degree is $(2,1)$. According to
Table~\ref{tab:Higgsdegree}, $H_a\bar{H}_b$ already has $(p,q)$ degree
$(2,0)$. Hence the moduli field $\phi_i$ must have degree $(0,1)$. In
eq.~\eqref{47} we found that only $4$ moduli $\phi_i$, $i=1,\dots,4$,
have the right $(p,q)$ degree. In other words, the majority of the
coefficients vanishes,
\begin{equation}
  \lambda_{iab}=0
  , 
  \qquad i=5, \dots, 19
  .
\end{equation}
In principle, the second selection rule imposes independent
constraints. It states that the total $[s,t]$ degree has to be
$[1,1]$. We showed that the allowed cubic terms $\phi_i H_a
\bar{H}_b$, $i=1,\dots,4$, all have the correct degree $[1,1]$.
Therefore, the $(p,q)$ and $[s,t]$ selection rule allow $\mu$-terms
involving $4$ out of the $19$ vector bundle moduli. Cubic terms
involving Higgs--Higgs conjugate fields and any of the remaining $15$
moduli are forbidden in the superpotential.

When the moduli develop non-zero vacuum expectation values these
superpotential terms generate Higgs $\mu$-terms of the form
\begin{equation} 
  \lambda_{iab} \, \langle \phi_{i}\rangle \, 
  H_{a} \bar{H}_{b}
  ,
  \qquad 
  i=1,\dots,4
  ,~
  a=1,2
  ,~
  b=1,2
  .
  \label{67-2}
\end{equation}
Moreover, the coefficient $\lambda_{iab}$ has no interpretation as an
intersection number, and therefore has no reason to be constant over
the moduli space. In general, we expect it to depend on the moduli. Of
course, to explicitly compute this function one needs the K\"ahler
potential which determines the correct normalization for all fields.

\section*{Acknowledgments}

This research was supported in part by the Department of Physics and
the Math/Physics Research Group at the University of Pennsylvania
under cooperative research agreement DE-FG02-95ER40893 with the
U.~S.~Department of Energy and an NSF Focused Research Grant
DMS0139799 for ``The Geometry of Superstrings.'' T.~P.~is partially
supported by an NSF grant DMS 0403884. Y.-H.~H.~is also indebted to
the FitzJames Fellowship of Merton College, Oxford.

\bibliographystyle{JHEP} \renewcommand{\refname}{Bibliography}
\addcontentsline{toc}{section}{Bibliography} \bibliography{muHH2}

\end{document}